# The Power of Fair Information Practices – A Control Agency Approach

**Christian Fernando Libaque-Saenz**
School of Engineering
Universidad del Pacífico
Lima, Peru
Email: cf.libaques@up.edu.pe

**Younghoon Chang**
Department of Computing & Information Systems
Sunway University
Selangor, Malaysia
Email: younghoonc@sunway.edu.my

**Siew Fan Wong**
Department of Computing & Information Systems
Sunway University
Selangor, Malaysia
Email: siewfanw@sunway.edu.my

**Hwansoo Lee**
IT Law Program, Graduate School
Dankook University
Yongin, Republic of Korea
Email: hanslee992@gmail.com

## Abstract

Most companies' new business practices are based on customer data. These practices have raised privacy concerns because of the associated risks. Privacy laws require companies to gain customer consent before using their information, which stands as the biggest roadblock to monetise this asset. Privacy literature suggests that reducing privacy concerns and building trust may increase individuals' intention to authorise the use of personal information. Fair information practices (FIPs) are potential means to achieve this goal. However, there is lack of empirical evidence on the mechanisms through which the FIPs affect privacy concerns and trust. This research argues that FIPs load individuals with control, which has been found to influence privacy concerns and trust level. We will use an experimental design methodology to conduct the study. The results are expected to have both theoretical and managerial implications.

## 1   INTRODUCTION

The development of information technologies (ITs) has brought significant changes to our lives. We use ITs for a wide range of activities, such as e-commerce, Internet banking, and social networking. The proliferation of smartphones and the mobile convenience of these devices have increased usage of related services. These human-computer interactions are producing huge and diverse amounts of customer data and thus we are creating digital footprints of our everyday activities (Girardin et al. 2008). Companies, on the other side, have started to heavily use and analyse these data together with other datasets to improve decision-making (Global Pulse 2012). For example, the Bloomberg Businessweek (2011)'s survey shows that 97 per cent of the respondents' companies (with more than $100-million revenues) have used some form of data analytics. Therefore, companies are expected to rely more on customer data in the future to reduce costs, develop new products and services, and stay competitive in the market (McAfee and Brynjolfsson 2012; McGuire et al. 2012).

The use of customer data, however, presents privacy challenges that should not be overlooked. ITs and the associated data collection activities have raised customers' privacy concerns. A recent study reported that Facebook users who committed 'virtual identity suicide' were mainly motivated by concerns over privacy (Woollaston 2013). Indeed, due to potential risks such as intrusion, public disclosure of embarrassing private facts, false light, or appropriation, individuals are uncomfortable with their data being collected. These uncertainties may lead individuals to stop using a service, or search for other companies that may provide similar service and protect their privacy at the same time (Dinev and Hart 2006, Liu et al. 2004).

Given the potential benefits and challenges of this trend involving customer data, the privacy domain has received considerable attention among researchers. Previous studies (see Table 1) have focused on individuals' intention to participate in activities involving the collection and use of their personal data, such as disclosing personal data and transacting on the Internet. In this study, we use the term 'privacy-related behavioural intention' to refer to these intentions. From an analysis of Table 1, the general conclusion is that privacy concerns have a negative impact on privacy-related behavioural intentions, while trust has a positive effect on these intentions. In other words, by reducing customers' privacy concerns and/or building their trust towards organisations, these institutions may monetise effectively on customer data.

| Author (Year) | Research Findings |
| --- | --- |
| Bart et al. (2005) | Trust has a significant positive effect on behavioural intention to disclose personal data to buy items through the Internet |
| Chellappa and Sin (2005) | Privacy concerns have a significant negative impact on behavioural intention to use online personalisation services, and trust has a significant positive effect on these intentions |
| Dinev and Hart (2006) | Behavioural intention to provide personal data to transact on the Internet is positively affected by trust, and negatively affected by privacy concerns |
| Gefen et al. (2003) | Trust has a significant positive effect on behavioural intention to provide personal data to use business-to-consumer (B2C) websites |
| Joinson et al. (2010) | Trust has a significant negative effect on intention to retain personal information (i.e., trust positively affects intention to disclose personal data) |
| Liu et al. (2004) | Trust positively affects behavioural intention to disclose personal data and transact on the Internet |
| Li (2011) | Users' privacy concerns negatively affect online privacy-related behavioural intentions such as disclosure of personal data |
| Malhotra et al. (2004) | Behavioural intention to disclose personal data is positively affected by trusting beliefs and negatively affected (indirectly) by privacy concerns |

*Table 1. Literature Review*

Culnan and Armstrong (1999)'s study on customers' willingness to be profiled when receiving personalised services sheds light on the potential effectiveness of fair information practices (FIPs) in reducing privacy concerns. FIPs are defined as companies' procedures to load customers with control over the disclosure and usage of their personal information (Culnan and Armstrong 1999). In the case of trust, Liu et al. (2004) conducted a study on individuals' intention to disclose personal data in e-commerce transactions and found similar results suggesting that FIPs may build trust. However, there is little empirical evidence on the effect of FIPs. For example, Culnan and Armstrong (1999) found that in situations where organisations use FIPs, the level of individual privacy concerns does not distinguish between individuals who are unwilling to be profiled and individuals who are willing to engage in this behaviour. Although the finding provides useful insights into the importance of FIPs, it is limited to the use of secondary data as proxy measurement of the variables. Liu et al. (2004) found a significant correlation between participants' perceptions of FIPs and trust. These findings, however, could not establish causality with certainty (Liu et al. 2004).

The objective of this study is to provide theoretical explanations on the role of FIPs in individuals' decisions in privacy-related behaviours, and to find empirical evidence for the causality of these relationships. Mobile applications are taken as the basis of this study because of their significant role in our lives (Hu et al. 2008), the dependence of their success on user adoption and continuance use of the platforms (Oghuma et al. 2015), and the importance of customer data to their business models (Dhar and Varshney 2011). It is expected that our results will provide companies with adequate tools to leverage on their access to customer data.

## 2 THEORETICAL BACKGROUND

This section includes the literature review on fair information practices (FIPs), information privacy, privacy concerns, trust theory, and the context of mobile applications.

### 2.1 Fair Information Practices (FIPs): A Control Agency Approach

Culnan and Armstrong (1999) defined FIPs as the procedures that load customers with control over the disclosure and usage of their personal information by data holders. FIPs have been created by the organisation for economic co-operation and development (OECD) in 1980. The US Federal Trade Commission (FTC) suggests that FIPs are made up of four main dimensions: notice, access, choice, and security (Liu et al. 2004). Notice refers to the practice of raising customers' awareness that personal data is being collected before its collection; access refers to the practice of proving customers with access to their collected personal data; choice refers to the practice of loading customers with power to decide which data about them could be used; and security refers to the practice of informing customers about the means for keeping their personal data secure (Libaque-Saenz et al. 2014, Liu et al. 2004). FIPs were a 'ready-made' tool rapidly adopted by countries. However, they remain an accepted 'black box.'

The control agency literature categorised control into self-control and proxy-control. The former is exerted by one's own self while the latter is exerted by powerful others (Xu et al. 2012). Yamaguchi (2001) defined proxy-control as individuals' attempts to align themselves with powerful others or powerful forces to gain control in situations where these individuals lack power to achieve their desired outcomes.

In the FIPs, the access and choice dimensions are based on self-control while the notice and security dimensions are based on proxy-control. The access dimension allows customers to exert direct control over their personal information by accessing data holders' databases to keep their data updated and free of errors. The choice dimension allows customers to exert direct control over their personal information by deciding which data could be used and for what purposes. For the notice dimension, customers could only rely on data holders' honesty to disclose clearly the ways in which they will use customer data (i.e., powerful force). For the security dimension, customers could only rely on data holders' capabilities to protect their data from security breaches (i.e., powerful others).

The privacy-trust literature suggests that FIPs may reduce customers' privacy concerns and build trusting beliefs (e.g., Culnan and Armstrong 1999; Liu et al. 2004). However, there is little empirical evidence on these effects. In addition, considering the number of recent security breaches such as the one happened in South Korea where personal data of 12 million customers of KT were stolen in 2014 after this network operator's website was compromised by hackers (Choi 2014), it is necessary to test the effectiveness of FIPs in an environment where people are aware of potential information risks. Prior research suggests that awareness of potential risks may increase privacy concerns and thus reduce individuals' positive expectations about the trustee — reduce trusting beliefs (Haggerty and Gazso 2002). Therefore, it is important to discuss theoretical explanations on the effect of the dimensions of FIPs on privacy concerns and trusting beliefs, and to find empirical evidence of these effects.

### 2.2 Information Privacy: Perceived Control

Westin (1967) defined information privacy as "the claim of individuals, groups, or institutions to determine for themselves when, how, and to what extent information about them is communicated to others." In other words, privacy is individuals' right to control the flow of their information. In a given situation, the extent to which others respect this right may vary. Therefore, prior research has extensively used 'perceptions' of control to operationalise privacy levels (e.g., Malhotra et al. 2004; Xu 2007; Xu et al. 2012). Perceived control is defined as individuals' beliefs of their ability to manage personal information (Xu et al. 2011). Prior research has found that loading customers with control over their personal data may mitigate their information privacy concerns (Brandimarte et al. 2012; Malhotra et al. 2004). When customers feel in control, their perceptions of possible opportunistic behaviour by trustees tend to decline, implying the development of customers' trusting beliefs (Brandimarte et al. 2012; Liu et al. 2004). In short, perceived control is a key variable in the privacy field and it is expected to directly affect privacy concerns and trust.

## 2.3 Information Privacy Concerns: Context-Specific

Past research has agreed with the conceptualisation of information privacy concerns as individuals worry about possible loss of control over personal data (Li 2011; Xu et al. 2011). This construct has been addressed from two perspectives: general and context-specific concerns (Li 2011). The privacy literature has underscored the importance of differentiating these two perspectives and highlighted that privacy concerns are more situation-specific than dispositional (Xu et al. 2011). Previous findings support the importance of context-specific over general information privacy concerns (e.g., Malhotra et al. 2004; Xu et al. 2012).

Prior findings have arrived at a general conclusion: information privacy concerns – general or specific – have a negative effect on privacy-related behavioural intentions (Chellappa and Sin 2005; Li 2011). For example, Chellappa and Sin (2005)'s work on the antecedents to the use of online personalised services found that privacy concerns have a negative impact on the likelihood of using these services. Li (2011) conducted an extensive literature review on online privacy studies and concluded that privacy concerns have a significant negative effect on the willingness to provide personal data for Internet transactions. Therefore, it is important to find empirical evidence supporting the effectiveness of various strategies in reducing customers' privacy concerns.

## 2.4 Trust Theory: Trusting Beliefs

The definition of trust varies across studies and disciplines. To conceptualise trust in our study, we will follow McKnight et al. (2002)'s proposed framework for initial trust, which refers to those relationships in which the actors are unfamiliar with each other. We adopt this framework because our experiment will be designed in a similar context. This framework segments the trust concept into three variables: 1) disposition to trust, which refers to the tendency to believe in the goodness of others; 2) institution-based trust, which refers to the beliefs that favourable conditions are in place to achieve a successful outcome; and 3) trusting beliefs, which are defined as the perceptions that the trustee has attributes that are beneficial to the trustor (McKnight et al. 2002).

Disposition to trust is developed gradually through lifelong socialisation (McKnight et al. 2002). Therefore, this construct remains unchangeable in situations where the actors interact for the first time (initial trust). Institution-based trust encompasses the factors that help to form an overall trust towards the trustee (McKnight et al. 2002). Trusting beliefs are cognitive evaluations that may be formed quickly, even before actors have meaningful information about each other (McKnight et al. 2002). In this study we focus on trusting beliefs to measure individuals' trust towards data holders. Trusting beliefs are made up of three dimensions: 1) competence, which refers to the trustors' beliefs that the trustees have the ability to perform to their expectations; 2) integrity, which refers to the trustors' beliefs that the trustees will be honest; and 3) benevolence, which refers to the trustors' beliefs that the trustees will not act in an opportunistic manner (Gefen et al. 2003).

Prior research has found empirical evidence on the positive impact of trusting beliefs on privacy-related behavioural intentions (Bart et al. 2005; Gefen et al. 2003). For example, Bart et al. (2005) conducted a research on the antecedents to individuals' intentions to buy items on the Internet, while Gefen et al. (2003) studied the antecedents to individuals' intentions to provide personal data to use a business-to-consumer (B2C) website. The results of both studies shed light on the significant positive effect of customers' trust towards the vendor on privacy-behavioural intentions. Hence, it is important to determine and empirically test the strategies for building customer trust.

## 2.5 Location-Based Mobile Applications: Setting the Context

Location-based mobile (LBM) applications (apps) are chosen to be the basis of this research because they depict well the studied phenomenon. First, these apps have gained importance in our daily lives. Indeed, LBM apps have become a prevalent phenomenon globally due to their ubiquitous context (Xu et al. 2012). LBM are mobile technologies that could be used anytime and anywhere (Sheng et al. 2008). Second, their success depends solely on users' intention to adopt and continue to use these services (Oghuma et al. 2015). If these apps lack of user traction, they will be unable to make profits and eventually disappear. Third, LBM apps, and mobile apps in general, are computer programs for smart devices that could be freely downloaded through platforms such as Apple App Store and Google Play. Therefore, these apps rely heavily on customer data because their business models are mostly based on personalised advertisements rather than subscriptions (Dhar and Varshney 2011). This situation may compromise customers' privacy. For example, Angwin and Valentino-Devries (2011) reported that Google and Apple apps have the capabilities to collect and track information about users' location automatically and that users do not have the choice to turn off this feature. This fact may raise users' concerns and reduce their perceptions toward providers' beneficial attributes.

## 3  RESEARCH HYPOTHESES

Our research is founded on the Control Agency Theory, which suggests that mechanisms triggering control perceptions over personal data can alleviate privacy concerns and build trusting beliefs (see Figure 1). We include paths from privacy concerns and trusting beliefs to behavioural intention for statistical testing only, given that previous studies have extensively discussed these effects. In this section, we present arguments for why each dimension of FIPs is expected to enhance control perception, which in turn may reduce privacy concerns and increase trusting beliefs.

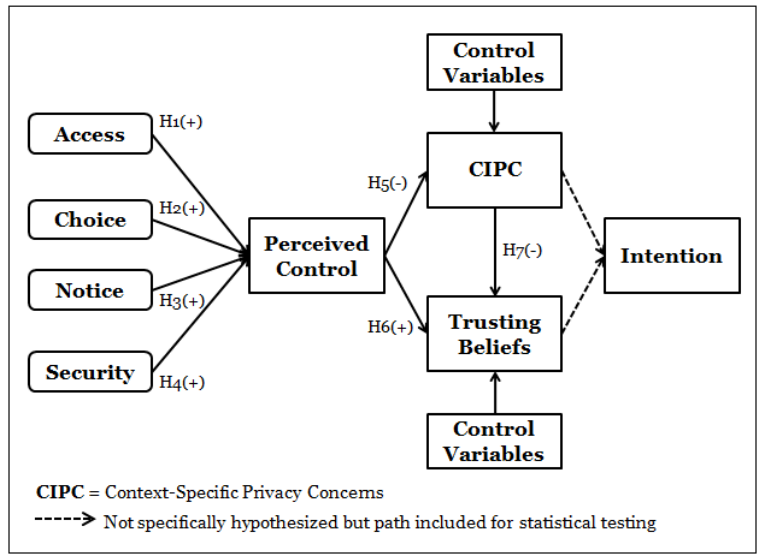

*Figure 1: Research Model*

### 3.1  Fair Information Practices (FIPs) and Perceived Control

As discussed in Section 2.1, the dimensions of FIPs are linked to the concept of control. In the case of the access dimension, customers can actively participate in controlling the quality of the stored data about them (data integrity). For example, users who are granted access to LBM apps' databases may maintain their location records to ensure the information is free of errors. Data integrity may raise individuals' perceptions of control (Suh and Han 2003). As for the choice dimension, customers can decide by themselves which information and in which ways the information can be used by data holders. This dimension allows customers to control the flow of their personal information. For example, users may decide which information about their location can be used through opt-in or opt-out alternatives installed on LBM apps' privacy tools, which may increase their control perceptions (Pearson 2009).

The notice dimension embraces the disclosure of the ways for collecting, using, and protecting customer data (privacy policies). These privacy policies load customers with knowledge to carry out informed decisions. Customers may be motivated to align themselves with this force to regulate and control the use of their personal information (Xu 2007, Xu et al. 2012). For example, users may rely on privacy policies to control the use of their data by LBM apps because these instruments constitute the rules of data usage.

The security dimension captures the means implemented by data holders to protect customer data (Pavlou and Fygenson 2006). Data holders represent powerful others because they are the agents of control in these situations, being those in charge of keeping customer data safe from security breaches. For instance, users of LBM apps may rely on the technologies implemented by providers to protect their personal information (e.g., privacy-enhancement technologies). Both notice and security dimensions lead us to the concept of proxy-control (execution of control through powerful forces or powerful others). It is important to notice that individuals' perceptions of proxy-control may increase their overall perceptual control (Bandura 2001).

Based on the above arguments, FIPs may increase customers' perceptual data control. Therefore, we hypothesise:

Hypothesis 1: Access positively affects perceived control

Hypothesis 2: Choice positively affects perceived control

Hypothesis 3: Notice positively affects perceived control

Hypothesis 4: Security positively affects perceived control

## 3.2   Perceived Control, Information Privacy Concerns, and Trusting Beliefs

This study focuses on situations where the actors interact for the first time. This type of scenarios is ruled by the actors' initial trust, which is influenced by the characteristics of the environment where they interact (McKnight et al. 2002). In our study, perceived control captures users' evaluations of the characteristics of different scenarios (i.e., the scenario with or without the FIPs).

In the case of the relationship between perceived control and information privacy concerns, Li (2011) concluded that individuals' evaluations of organisational environmental factors (e.g., enforce privacy policies) may alleviate their privacy concerns. Therefore, perceived control (construct capturing environmental factors) is expected to exert impact on privacy concerns. In fact, the control literature argues that perceptual control over data may mitigate customers' privacy concerns because customers are motivated to expect positive outcomes in high-control situations (Brandimarte et al. 2012; Malhotra et al. 2004). The psychology literature also suggests that a lack of control may lead customers to feel anxiety, worries, or concerns (for a review, see Rapee et al. 1996). Consequently, we argue that perceived control has a negative impact on information privacy concerns, which is consistent with previous studies (Malhotra et al. 2004; Xu et al. 2011).

Hypothesis 5: Perceived control negatively affects information privacy concerns

The discussion on individuals' perceptions of environmental characteristics leads us to the concept of institution-based trust because this construct is defined as "the belief that needed structural conditions are present […] to enhance the probability of achieving a successful outcome" (McKnight et al. 2002). As discussed in Section 2.1, perceived control gauges the presence of conditions such access, choice, notice, and security, which may enhance the likelihood to achieve positive outcomes. Accordingly, perceived control represents institution-based trust. McKnight et al. (2002) postulates that institution-based trust has a positive impact on trusting beliefs because individuals tend to perceive that actors have trustworthy attributes in 'save' situations. Therefore, perceived control has a positive impact on trusting beliefs (Das and Teng 2001; Joinson et al. 2010). We hypothesize that a lack of control may increase customers' perceptions toward a potential opportunistic behaviour by data holders and thus reduce their trusting beliefs towards these organisations.

Hypothesis 6: Perceived control positively affects trusting beliefs

We also hypothesise a negative impact of privacy concerns on trusting beliefs as follows. First, individuals who are high on privacy concerns may be sceptical about the capabilities of data holders to efficiently manage their information privacy (Slyke et al. 2006). Second, for individuals who are highly concerned about their privacy, potential business opportunities based on customer data may exacerbate their perceptions of potential dishonest and opportunistic data-holder behaviours (Slyke et al. 2006). This hypothesis is consistent with past findings (Malhotra et al. 2004; Van Slyke et al. 2006).

Hypothesis 7: Information privacy concerns negatively affect trusting beliefs

Our research model implies that the effect of the FIPs on privacy concerns and trusting beliefs is fully mediated by individuals' perceptual control. We will empirically test this proposed indirect effect.

## 3.3   Control Variables

Xu et al. (2012) suggests that privacy concerns could be influenced by three factors: 1) personal characteristics, 2) situational cues based on which individuals assess the consequences of engaging in privacy-related behaviours, and 3) interpersonal differences. The first category embraces demographics. Our study have included age, gender, and education as control variables because individuals who are less likely to be concerned about their privacy are more likely to be male, young, and less educated (Culnan 1995, Sheehan 1999). In the second category we included disposition to trust and desire for control. The former is defined as individuals' tendency to believe in the goodness of others (McKnight et al. 2002), while the latter refers to individuals' tendency to desire information control (Phelps et al. 2000). Disposition to trust has a positive effect on trusting beliefs (McKnight et al. 2002), whereas individuals who desire greater information control tend to have higher levels of privacy concerns (Phelps et al. 2000). Finally, in the third category we included prior experience of privacy invasion and awareness of potential privacy breaches. Smith et al. (1996)' findings suggest that individuals who have experienced privacy invasion are more likely to have high privacy concerns than those who have not been victims of these torts. Similarly, previous research suggests that the more knowledgeable people become about potential privacy breaches, the higher are their privacy concerns (Haggerty and Gazso 2002).

In sum, we argue that these variables may have a potential impact on both privacy concerns and trusting beliefs (variables closely related). Therefore, we control for their influences to isolate the effect of FIPs on both privacy concerns and trusting beliefs.

## 4 RESEARCH METHODOLOGY

Given that this study constitutes a research-in-progress work, this section describes the proposed procedures for data collection and future data analyses.

### 4.1 Research Design and Experimental Manipulations

An experimental design will be used because it is appropriate for testing causality (Sheng et al. 2008). The presence or absence of each of the dimensions of FIPs will be manipulated independently, and their effect on consumers' perceived control, privacy concerns, and trusting beliefs will be analysed. The experiment will be made up of 2 (with/without access) X 2 (with/without choice) X 2 (with/without notice) X 2 (with/without security) scenarios based on a between-subject-factorial design. Following prior literature, these four factors will be operationalised using a scenario-based methodology (Camponovo et al. 2004).

This research focuses on a mobile application (app) that provides paperless discounts. This app is a platform between merchants and customers for promotions and discounts about products and services. The app is connected to the databases of mobile telephone service providers and is capable of automatically accessing customer location at any time to send information about discounts in real-time. Sixteen experimental scenarios will be created. The presence of each of the dimensions of FIPs will be operationalised by explicitly telling the participants the followings:

- for the access dimension, 'they are able to access their past shopping records and correct them if any mistake exists;'
- for the choice dimension, 'they are able to give or revoke their consent to use any personal data stored by the app;'
- for the notice dimension, 'they will be able to revise the app's practices to use their personal information (i.e., how long their personal information will be stored, how their personal information will be used, and with whom their personal information will be shared);' and
- for the security dimension, 'this app is using data encryption technology, which is used to keep information safe from security breaches.'

The absence of the dimensions of FIPs will be operationalised by not including any of these statements in the description of the scenarios.

### 4.2 Measurement Instrument

Behavioural intention are measured with 3 items from Sheng et al. (2008), privacy concerns with 4 items from Xu et al. (2011), trusting beliefs with 3 items from Pavlou and Fygenson (2006), while perceived control, trust disposition, desire for control, previous experience, and awareness of potential risks are measured with 5, 3, 3, 1, and 1 items respectively from Xu et al. (2012). All measurement items are drawn from the literature and adapted to our research context. The items are anchored on a 7-point Likert scale. All items are reflective. However, the dimensions of FIPs are dichotomous variables, which are formative in nature (Hair et al. 2013).

### 4.3 Research Procedures and Data Collection (Future Step)

The scenarios will be administrated through questionnaires made up of three sections: 1) items related to the control variables, 2) a randomly assigned scenario (each participant will have an equal and independent chance to be assigned to any of the sixteen scenarios), and 3) items related to demographics. In the second section, participants will be told that Company A will introduce a new mobile app and consumers' feedback would help to evaluate it. Then, the description of one of the sixteen scenarios will be randomly presented followed by four manipulation check questions — one question for each dimension of FIPs — and items related to participants' perceptions of control, privacy concerns, and trusting beliefs.

The sample will be chosen from USA consumers and collected through Amazon Mechanical Turk, which is a web-based platform where employers (called requesters) post outsourced tasks for an anonymous network of labourers (called workers) who receive compensation for their contribution (Steelman et al. 2014). This web site has been shown to be effective in data collection (Steelman et al. 2014). We plan to conduct pilot tests to ensure that our scenarios are able to detect differences. After checking the appropriateness of our experiment, and the reliability and validity of our measurement items, we plan to collect the full-scale sample. To determine the minimum sample size we follow Chin (1998)'s rule-of-thumb which establishes that a sample size should be at least 10 times (1) the largest number of formative indicators or (2) the largest number of independent variables impacting a dependent variable, whichever is greater. In our research model the largest number of formative indicators is 1 (for access, choice, notice and security), whereas the largest number of independent variables affecting a dependent variable is 4 (for perceived control). We expect to collect 500 usable responses, which is more than adequate for the PLS estimation procedures (minimum sample size is 40 participants). Responses from the participants who fail to correctly answer the manipulation check questions will be removed from data analysis.

## 5   DATA ANALYSIS (FUTURE STEP)

For data analysis, we will use partial least squares (PLS) technique for two reasons: 1) our study is exploratory in nature and is in an early stage of theory development (Komiak and Benbasat 2006), and 2) we use formative items to model the dimensions of FIPs, which are dichotomous variables (Hair et al. 2013). SmartPLS will be used as analysis tool (Ringle et al. 2005).

A two-step approach will be used to check the experimental manipulations, ensure a successful random assignment of participants to scenarios, and assure that differences in perceptual control are indeed due to the dimensions of FIPs. First, the conditions about the existence or absence of each dimension of FIPs will be checked through yes/no questions to confirm that the participants understand the scenarios. Second, statistical tests (e.g., chi-square, ANOVA) will be conducted to examine the differences in control variables across scenarios.

As for the measurement model, reliability and convergent validity will be assessed through three tests: 1) reliability of items, 2) internal consistency, and 3) average variance extracted (AVE). On the other hand, discriminant validity will be checked by comparing the square root of the AVEs with the correlations among the constructs.

In addition, we will check the absence of multicollinearity among variables and common method bias. The former will be performed by examining that variance inflation factor (VIF) values are less than 5 when performing regression analysis (Hair et al. 2011). As for the latter, we will check that no single factor accounts for the majority of the variance by performing the Harman's single-factor test (Hazen et al. 2011).

To assess the explanatory power of our structural model, we will analyse the amount of variance explained in the endogenous constructs ($R^2$) and paths between variables. We will also compare three models using Cohen (1988)'s method ($f^2$). Model 1 represents the full model including control variables, Model 2 represents the effect of our theoretical constructs (excluding control variables), and Model 3 represents only the effect of the control variables, which will serve as the basis to analyse the impact of the theoretical constructs.

Finally, we will assess the mediation role of perceptual control in the relationship between FIPs and both privacy concerns and trusting beliefs. For this task, Baron and Kenny (1986) and Sobel (1982)'s tests will be used.

## 6   IMPLICATIONS (EXPECTED)

This section discusses the expected implications of this study's future findings for both theory and practice.

### 6.1 For Theory

The present research attempts to provide a discussion on the importance of fair information practices (FIPs) in privacy-related behaviours. By analysing each of the dimensions of FIPs, this study may provide clear insights into self-control and proxy-control mechanisms for addressing individuals' privacy concerns and their lack of trusting beliefs.

Accordingly, this study's findings are expected to propose a theoretical background to explain the effect of FIPs on control, which in turn affects privacy concerns and trust. The results may demonstrate that FIPs load individuals with perceptual control. Also, this study may empirically demonstrate the importance of using FIPs in strategies to reduce privacy concerns and building trust towards organisations.

Finally, a comparison among the effects of the dimensions of FIPs on perceived control may shed light on individuals' preferences for exerting control in situations implying the collection and use of their personal information.

### 6.2 For Practice

In spite of the potential business benefits related to customer data usage, most companies may be missing out on these opportunities because they lack of effective tools to encourage their customers to engage in privacy-related behaviours. In terms of the FIPs, Schwaig et al. (2006)'s work shows that most of the Fortune-500 firms only complied with the notice dimension and failed to efficiently address the other dimensions. Our study attempts to provide companies with a better understanding on the role of the FIPs in reducing privacy concerns and building trust, which may help them to redesign their business strategies in this area. With the FIPs, customers may be more willing to disclose and share personal information with companies. However, companies should carefully leverage on FIPs and be honest with their capabilities to comply.